\documentclass[pra,aps,twocolumn,superscriptaddress,floatfix,noshowpacs,10pt]{revtex4-2}

\usepackage[utf8]{inputenc}
\usepackage{microtype}
\usepackage{amsmath,amssymb}
\usepackage{graphicx}
\usepackage[colorlinks=true,citecolor=blue]{hyperref}
\usepackage{cases}

\usepackage{bm}

\newcommand{\bs}[1]{{\boldsymbol{#1}}}

\newcommand{\br}{\bs{r}}

\newcommand{\bq}{\bs{q}}

\begin{document}

\title{Dynamical phase transition of light in time-varying nonlinear dispersive media} 

\author{Nicolas Cherroret}
\email{nicolas.cherroret@lkb.upmc.fr}
\affiliation{Laboratoire Kastler Brossel,
Sorbonne Universit\'{e}, CNRS, ENS-PSL Research University, 
Coll\`{e}ge de France; 4 Place Jussieu, 75005 Paris, France}

\begin{abstract}
We demonstrate the existence of a prethermal dynamical phase transition (DPT) for fluctuating optical beams propagating in nonlinear dispersive  media. The DPT can be probed by suddenly changing in time the dispersion and nonlinearity parameters of the medium (thus realizing a ``temporal interface''), a procedure that emulates a quench in a massive $\varphi^4$ model.
Above a critical value of the quench identifying the transition, the fluctuating beam after the temporal interface is characterized by a correlation length that diverges algebraically at the transition. Below the critical quench, the beam exhibits an algebraic relaxation and a self-similar scaling. Our analysis also reveals a dimensional cross-over of the critical exponent, a characteristic feature of the optical DPT. 
\end{abstract}

\maketitle

\section{Introduction}

In the non-equilibrium physics of isolated many-body systems, dynamical phase transitions (DPTs) have recently sparked considerable interest, as examples of critical phenomena characterized by scaling properties different from their equilibrium counterparts. 
Loosely speaking, a DPT is associated with the emergence of well distinct temporal evolutions of certain observables  following a quantum quench. From this general definition, however, several qualitatively different types of DPTs have been identified. A first type, for instance, arises in the time evolution of Loschmidt echos, which may exhibit a cusp at a critical time upon quenching a parameter of the Hamiltonian, with the rate function of the echo vanishing at the critical time \cite{Jurcevic2017, Weitenberg2018, Heyl2018}. 
A second category of non-equilibrium critical phenomena has been observed for strong cooling quenches of three-dimensional quantum gases. Following the quench, the momentum distribution of the gas exhibits a universal, spatio-temporal self-similar scaling governed by a set of dynamical exponents \cite{Berges2008, Nowak2011, Prufer2018, Erne2018, Eigen2018}. A characteristic feature of this phenomenon, dubbed non-thermal fixed point, is to be governed by the collisions between the quasi-particle excitations of the cold gas \cite{Schmied2019}. 
A third type of DPT finally, which is the object of the present paper, arises in the so-called \emph{prethermal} regime of many-body systems, where the quasi-particle collisions are, in contrast, mostly ineffective \cite{Marino2022}. Prethermalization refers to an intermediate regime of times following a quench, where the dynamics is governed by  excitations whose properties are renormalized by interactions but which can be considered independent. A prethermal dynamics naturally shows up,  in particular,  in weakly-interacting systems close to integrability \cite{Berges2004, Mori2018, Larre2018, Martone2018, Mallayya2019, Bardon-brun2020}, as recently observed experimentally  in cold atom \cite{Gring2012, Langen2013, Tang2018} and photonic \cite{Abuzarli2022} setups.
In that context, a prethermal DPT corresponds to the emergence of qualitatively different dynamics of the system's correlations when quenching a control parameter of the Hamiltonian around a critical value \cite{Marino2022}. Theoretically, prethermal  DPTs have been especially described in fully-connected spin models \cite{Das2006, Sciolla2011, Defenu2018, Lerose2019} and in $\varphi^4$ field theories with $O(N)$ symmetry in the large $N$ limit \cite{Chandran2013, Sciolla2013, Smacchia2015, Chiocchetta2017, Maraga2015, Mitra2015, Mitra2016, Halimeh2021}.  On the experimental side, observations and characterizations of prethermal DPTs  have been achieved with cold atoms, among which trapped ions \cite{Zhang2017}, atoms in optical cavities \cite{Muniz2020}, Fermi gases \cite{Smale2019} and spinor condensates \cite{Yang2019, Tian2020}.

In this paper, we theoretically demonstrate  the existence of a prethermal dynamical phase transition in a closed optical system made of a fluctuating light beam propagating in a nonlinear, dispersive dielectric medium. In the last decades, such optical platforms have been extensively investigated due to their ability to emulate with light the low-energy physics of quantum gases \cite{Carusotto2013}. In particular, laser beams propagating in nonlinear atomic vapors \cite{Glorieux2023} have proven to constitute a flexible tool to explore non-equilibrium phenomena such as thermalization \cite{Sun2012, Santic2018}, prethermalization and light-cone spreading \cite{Abuzarli2022}, Zakharov-type oscillations \cite{Steinhauer2022}, vortex dynamics \cite{Azam2022}, parametric resonances \cite{Martone2023} or turbulence \cite{Abobaker2023}. Likewise, in optical fibers the interplay between dispersion and nonlinearity has revealed interesting prethermal effects such as the Fermi-Pasta-Ulam-Tsingou recurrences \cite{Simaeys2001, Mussot2014} and an associated mechanism of broken symmetry \cite{Mussot2018}. 

The existence of an optical DPT discussed in the present work relies on the close resemblance between the wave equation governing light propagation in nonlinear dispersive media and the equation of motion of a massive, classical $\varphi^4$ field theory. From this observation, we propose an optical quench protocol allowing to trigger such a transition, based on a temporal change of the dispersion parameters of the medium (Sec. \ref{Sec:model}), and we identify the precise condition under which the transition effectively occurs (Sec. \ref{Sec:DPTgeneral}). In Sec. \ref{Sec:above} we then characterize the postquench dynamics of the optical beam above the transition point. This analysis, in particular, reveals the existence of \emph{two} critical exponents characterizing the transition. In the close vicinity of the critical point first, the critical exponent $\nu$ coincides with that of the equilibrium quantum phase transition of the underlying two-dimensional $\varphi^4$ theory. When moving away  from the transition, however, we find that $\nu$ crosses-over to the value expected for an equilibrium quantum phase transition in dimension 3. This dimensional cross-over is a characteristic feature of the optical DPT. Below the transition point, the dynamics exhibits scale invariance and self-similar scaling, which we describe both theoretically and analytically in Sec. \ref{Sec:below}, following previous ideas developed in the context of the $O(N)$ model \cite{Chandran2013, Sciolla2013, Smacchia2015, Chiocchetta2017, Maraga2015, Mitra2015, Mitra2016, Halimeh2021}. At last, in Sec. \ref{Sec:vapor}, we particularize the problem to a concrete system, a light beam propagating in a resonant atomic vapor, and deduce a possible phase diagram of the DPT in that type of medium. Sec. \ref{Sec:conclusion} finally concludes the paper.

\section{The model}
\label{Sec:model}

\subsection{Wave equation in nonlinear dispersive media}

We consider an optical  beam propagating in a dielectric medium in which the electric field is governed by the Helmholtz equation 
\begin{equation}
\label{eq:Helmholtz}
\nabla\!\times\![\nabla\!\times\! \textbf{E}(\br,\omega)]=\frac{\omega^2}{c^2}\epsilon(\omega,\br)\textbf{E}(\br,\omega),
\end{equation}
where the relative permittivity $\epsilon(\omega,\br)=\epsilon_\text{L}(\omega)+\epsilon_\text{NL}(\br)$ decomposes into a linear, dispersive part $\epsilon_\text{L}(\omega)$, and a nonlinear part $\epsilon_\text{NL}(\br)\propto |\textbf{E}|^2$ that depends quadratically on the wave field (Kerr effect). We further suppose that the beam is mostly directed along the axis $z$, and that its spectrum is centered around a carrier frequency $\omega_0$. This invites us to express the electric field as
\begin{equation}
\label{Eparaxial}
\textbf{E}(\br,t)=\text{Re}\left[\boldsymbol{\mathcal{E}}(\br_\perp,z,t)e^{i(k_0z-\omega_0t)}\right],
\end{equation}
where we isolated the envelope $\boldsymbol{\mathcal{E}}(\br_\perp,z,t)$, assumed to be a slowly-varying function of the transverse, $\br_\perp\equiv(x,y)$, longitudinal, $z$, and temporal, $t$, coordinates. From now on, we assume that this envelope has a fixed polarization in the $(x,y)$ plane, and we focus on the evolution of the corresponding amplitude, denoted by $\mathcal{E}$. 
In Eq. (\ref{Eparaxial}), we also introduced the optical wave number $\smash{k_0\equiv \sqrt{\epsilon_\text{L}(\omega_0)}\omega_0/c}$ at the carrier frequency, with $c$ the speed of light in vacuum.

To account for the dispersion of the medium, we Taylor expand the linear part of the squared wave vector in the right-hand side (r.h.s.) of Eq. (\ref{eq:Helmholtz}) around $\omega_0$ (slowly-varying envelope approximation) \cite{Rosanov2002}:
\begin{equation}
\label{eq:taylork2}
k^2(\omega)\equiv \frac{\omega^2}{c^2}\epsilon_\text{L}(\omega)\simeq k_0^2+\frac{2k_0}{v}(\omega-\omega_0)+D(\omega-\omega_0)^2,
\end{equation}
where $v\equiv (\partial k/\partial\omega)^{-1}$ is the group velocity and $D\equiv (1/2)(\partial^2 k^2/\partial\omega^2)$ is the quadratic dispersion parameter, which we assume positive from now on. Inserting Eqs. (\ref{Eparaxial}) and (\ref{eq:taylork2}) into Eq. (\ref{eq:Helmholtz}) and dropping terms involving second-order derivatives with respect to $z$ (paraxial approximation), we find
\begin{equation}
\left(D\partial^2_t\!-\! \Delta_\perp\!-\!2ik_0\partial_z\!-\!\frac{2i k_0}{v}\partial_t\!-\!\frac{\omega_0^2}{c^2}\epsilon_\text{NL}\right)\mathcal{E}=0,
\end{equation}
where $\Delta_\perp$ is the Laplace operator in the transverse plane $(x,y)$. As a last step, we introduce the new field variable
\begin{equation}
\label{phi_vs_E}
\phi(\br_\perp,z,t)\equiv\mathcal{E}(\br_\perp,z,t)e^{-i\mu t}
\end{equation}
where $\mu\equiv k_0/(D v)$. In this frame, the wave equation becomes  
\begin{equation}
\label{final_wave_eq}
\left(D\partial^2_t\!-\!\Delta_\perp\!-\!2i k\partial_z\!+\!\frac{k_0^2}{Dv^2}\!+g|\phi|^2\right)\phi(\br_\perp,z,t)\!=\!0,
\end{equation}
where 
we defined the nonlinear parameter $g$ such that $-(\omega_0^2/c^2)\epsilon_\text{NL}\equiv g |\phi|^2$. In the following, we suppose $g>0$, corresponding to a defocusing nonlinearity.


At this stage, it is already interesting to notice that the wave equation (\ref{final_wave_eq}) resembles  a nonlinear Klein-Gordon-type equation, i.e., the  equation of motion of a classical, massive $\varphi^4$ theory with coupling constant $g$ and ``mass'' $k_0^2/(Dv^2)$. 
Generally speaking, the $\varphi^4$ field theory is a prototypical model  describing the large-scale behavior of a broad range of systems near a second-order equilibrium phase transition separating a ``disordered'' and an ``ordered'' phase \cite{Kleinert2001, Moshe2003}. In the present context though, the propagating wave is a priori not in a state of thermal equilibrium, so that probing such a phase transition with light is not obvious. The $\varphi^4$ model, nevertheless, is also known to host a \emph{dynamical} phase transition \cite{Chandran2013, Sciolla2013, Marino2022}, which does not necessarily require to start from a thermal state. Such a DPT usually arises when performing a  \emph{temporal} change (quench) of a control parameter --often the mass-- around a certain critical value. Following the quench, the system exhibits distinctive dynamical evolutions on each side of this critical point.
In the next sections, we give consistency to this discussion by  introducing a physical protocol allowing to probe a DPT of that type with light governed by the wave equation (\ref{final_wave_eq}).

\subsection{Prequench optical state and quench protocol}

\begin{figure}
\includegraphics[scale=0.65]{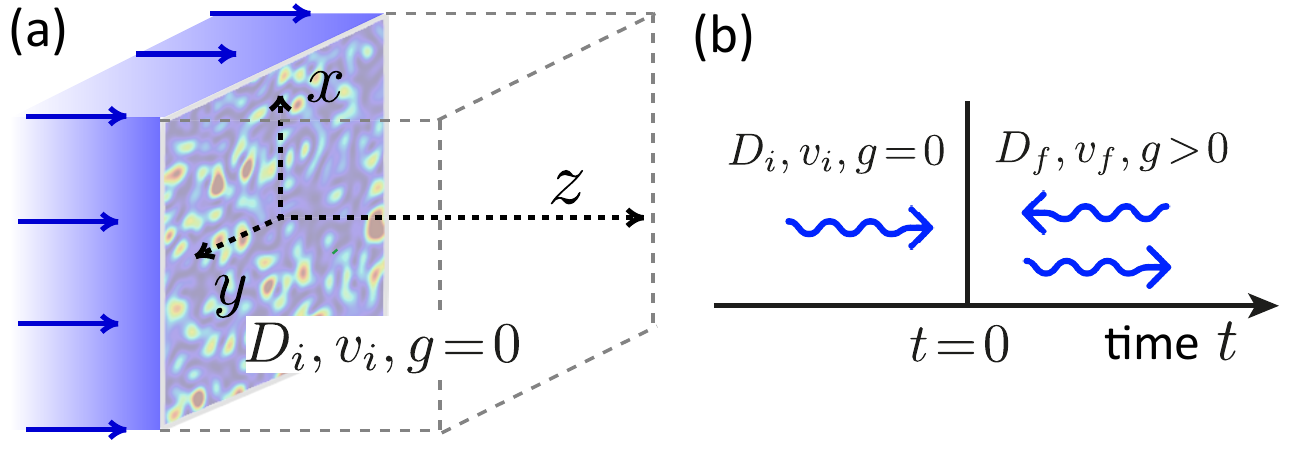}
\caption{
(a) We consider an optical beam propagating for $t<0$ in a dispersive dielectric medium with dispersion parameters $D_i$ and $v_i$ and no nonlinearity. The wave is assumed to exhibit  spatial fluctuations in the $(x,y)$ plane, as well as frequency fluctuations (not shown in the figure). (b) At $t=0$, we suppose that the dispersion and nonlinearity parameters are suddenly quenched from $(D_i,v_i,g=0)$ to $(D_f,v_f,g>0)$. This defines a temporal interface, beyond which two waves propagating forward and backward in time emerge.}
\label{fig:setup}
\end{figure}

To probe the optical DPT behind Eq. (\ref{final_wave_eq}), we propose to consider a \emph{fluctuating} beam initially propagating for $t<0$ in a dispersive medium with $v=v_i$, $D=D_i>0$ and no nonlinearity ($g=0$), see Fig. \ref{fig:setup}(a).
Defining the Fourier transform $\phi(\bq_\perp,z,\omega)\equiv \int d^2\br_\perp\int dt\phi(\br_\perp,z,t)e^{-i\bq_\perp\cdot\br_\perp+i\omega t}$, we  assume that the spectrum of the beam at $z=0$ takes the form 
\begin{equation}
\label{eq:spectrum}
\phi^{(0)}(\bq_\perp,\omega)\equiv \phi(\bq_\perp,z=0,\omega)=\sqrt{I}\, \phi_T(\bq_\perp)\phi_F(\omega),
\end{equation}
where $I$ is the beam intensity, and $\phi_T$ and $\phi_F$ are random fields which respectively encode spatial fluctuations in the transverse plane and frequency fluctuations. In practice, the spatial fluctuations can be obtained by, e.g., imprinting a speckle pattern onto the wavefront of a laser \cite{Abuzarli2022}, while the frequency fluctuations are associated  with deviations of the beam from pure monochromaticity.
As a first property of these fluctuations, we impose that the statistical average of the field $\smash{\phi^{(0)}}$ vanishes, which is  for instance realized by setting
\begin{equation}
\label{eq:mean_field}
\langle\phi_T(\bq_\perp)\rangle=0.
\end{equation}
Defined in this way, the optical field for $t<0$ can be seen as a kind of optical analogue to a thermal  system belonging to a disordered phase. We further suppose that the spatial and frequency fluctuations are  translation-invariant and stationary, respectively, i.e., their
two-point correlators obey  
\begin{equation}
\label{correlator_q}
\langle\phi_T^*(\bq_\perp)\phi_T(\bq_\perp')\rangle\!=\!\delta^2(\bq_\perp\!-\!\bq_\perp')S_T(\bq_\perp)
\end{equation}
and 
\begin{equation}
\label{correlator_omega}
\langle\phi_T^*(\omega)\phi_T(\omega')\rangle\!=\!\delta(\omega\!-\!\omega')S_F(\omega).
\end{equation}
These relations also define the spatial and frequency fluctuation spectra, $S_T(\bq_\perp)$ and $S_F(\omega)$, which we choose normalized: $\int d^2\bq_\perp/(2\pi)^2S_T(\bq_\perp)=\int d\omega/(2\pi) S_F(\omega)=1$. In the quench protocol presented below, we will see that the spatial spectrum can be traced out of the description, such that there is no need to specify its exact shape at this stage. As for the frequency spectrum, in the following we consider the simple Lorentzian shape
\begin{equation}
\label{eq:Lorenztian}
S_F(\omega)=\frac{2\gamma}{(\omega-\mu)^2+\gamma^2}
\end{equation}
of bandwidth $\gamma$. Notice that $S_F(\omega)$ is centered around $\omega=\mu$, which does correspond  to a spectrum of the original field $\textbf{E}$ centered around the carrier frequency $\omega_0$ by virtue of the definitions (\ref{Eparaxial}) and (\ref{phi_vs_E}).

To trigger a DPT in that system, the key idea is to perform a temporal change of the optical parameters that emulates a quench from the disordered to the ordered phase in the underlying $\varphi^4$ theory \cite{Marino2022}. To achieve this goal, we assume that, at $t=0$, the dispersion parameters $(v,D)$ and the nonlinear strength $g$ of the medium are suddenly changed according to the following protocol:
\begin{equation}
\label{eq:quench_protocol}
(D_i>0,v_i,g=0)\to (D_f>0,v_f,g>0),
\end{equation}
where $D_f$ and $v_f$ can, at this stage, a priori take arbitrary positive values. As a remark, from the physical point of view this quench protocol defines a ``time-varying'' interface in the dielectric medium at $t=0$, as illustrated in Fig. \ref{fig:setup}(b). This interface generically gives rise, for $t>0$, to two waves propagating forward and backward in time, analogously to the well known transmitted and reflected waves arising at a spatial interface between two dielectric media \cite{Pendry2022}. 
We will come back more quantitatively to this interpretation at the end of Sec. \ref{Sec:xi}.

Given Eq. (\ref{eq:spectrum}), the wave field in the prequench regime $t<0$, solution of the wave equation (\ref{final_wave_eq}), explicitly reads
\begin{align}
\label{phi_prequench}
\phi(\br_\perp,z,t<0)\!&=\!\int\! \!\! \int\frac{d\omega}{2\pi} \frac{d^2\bq_\perp}{(2\pi)^2}
\phi(\bq_\perp,0,\omega)e^{i\bq_\perp\cdot\br_\perp-i\omega t}
\nonumber\\
&\times\exp\Big[\!-\frac{iz}{2k}\Big(\bq_\perp^2\!-\!D_i\omega^2\!+\frac{k_0^2}{D_i v_i^2}\Big)\Big].
\end{align}
Starting from this optical state and from the quench protocol (\ref{eq:quench_protocol}), in the next section we examine the dynamical evolution of the wave field for $t>0$.

\section{Dynamical phase transition of light}
\label{Sec:DPTgeneral}

\subsection{Mean-field postquench dynamics}
\label{Sec:HFB}
To find the solution of  the wave equation (\ref{final_wave_eq}) for $t>0$, we use the Ansatz
\begin{align}
\label{eq:Ansatz}
\phi(\br_\perp,z,t>0)\!&=\!\int\! \!\! \int\frac{d\omega}{2\pi} \frac{d^2\bq_\perp}{(2\pi)^2}
\phi^{(0)}(\bq_\perp,\omega)e^{i\bq_\perp\cdot\br_\perp} f_\omega(t)
\nonumber\\
&\times\exp\Big[\!-\frac{iz}{2k}\Big(\bq_\perp^2\!-\!D_i\omega^2\!+\frac{k_0^2}{D_i v_i^2}\Big)\Big],
\end{align}
which has essentially the same form as Eq. (\ref{phi_prequench}), except for the unknown function $f_\omega(t)$. The latter can be found by imposing  that Eq. (\ref{eq:Ansatz}) is solution of the wave equation (\ref{final_wave_eq}) with $D=D_f$, $v=v_f$ and $g\ne0$.
This yields
\begin{align}
\label{modal_waveeq}
&\int\! \!\! \int\frac{d\omega}{2\pi} \frac{d^2\bq_\perp}{(2\pi)^2}\phi^{(0)}(\bq_\perp,\omega)e^{i\bq_\perp\cdot\br_\perp-\frac{iz}{2k_0}(\bq_\perp^2\!-\!D_i\omega^2\!+\frac{k_0^2}{D_i v_i^2})}\\
&\!\times\!\Big[D_f\ddot f_\omega (t)\!+\!\Big(D_i\omega^2\!+\!\frac{k_0^2}{D_f v_f^2}\!-\!\frac{k_0^2}{D_i v_i^2})f_\omega(t)\Big]\!+\!g|\phi|^2\phi\!=\!0.\nonumber
\end{align}
When evaluated with the Ansatz (\ref{eq:Ansatz}), the last nonlinear term $g|\phi|^2\phi$ in the left-hand side involves products of three random fields $\phi^{(0)}$ at different momenta and frequencies. To simplify it, we employ a mean-field Hartree-Fock-Bogoliubov approximation, a truncation scheme that consists in neglecting all correlation functions beyond the second one \cite{Blagoev2001}. While this scheme cannot capture the long time evolution after the quench, it is known to accurately describe the intermediate time scales, i.e. the prethermal dynamics, where a DPT is expected to take place.
In the term $g|\phi|^2\phi$, this approximation amounts to applying the factorization rule:
\begin{align}
\phi^{(0)*}(\bq_1,\omega_1)&\phi^{(0)}(\bq_2,\omega_2)\phi^{(0)}(\bq_3,\omega_3)\to\nonumber\\
&\langle\phi^{(0)*}(\bq_1,\omega_1)\phi^{(0)}(\bq_2,\omega_2)\rangle\phi^{(0)}(\bq_3,\omega_3)\nonumber\\
+&\langle\phi^{(0)*}(\bq_1,\omega_1)\phi^{(0)}(\bq_3,\omega_3)\rangle\phi^{(0)}(\bq_2,\omega_2).
\end{align}
Making use of Eqs. (\ref{correlator_q}) and (\ref{correlator_omega}) and invoking the normalization condition for $S_T(\bq_\perp)$, 
we infer:
\begin{align}
g|\phi|^2\phi&\simeq2 g I \!\int\! \!\! \int\ \frac{d\omega}{2\pi} \frac{d^2\bq_\perp}{(2\pi)^2}\phi^{(0)}(\bq_\perp,\omega)\\
&\times\int \frac{d\omega'}{2\pi}S_F(\omega')|f_{\omega'}(t)|^2e^{i\bq_\perp\cdot\br_\perp-\frac{iz}{2k_0}(\bq_\perp^2\!-\!D_i\omega^2\!+\frac{k_0^2}{D_iv_i^2})}\nonumber.
\end{align}
Inserting this relation into Eq. (\ref{modal_waveeq}), we  finally obtain a closed equation for $f_\omega(t)$:
\begin{equation}
\label{fomega_eq}
\ddot f_\omega (t)\!+\!\Big[\frac{D_i}{D_f}\omega^2\!+\!m_\text{eff}(t)\Big]f_\omega(t)=0,
\end{equation}
where
\begin{equation}
\label{meff_def}
m_\text{eff}(t)\equiv  m+\frac{2gI}{D_f}\int_{-\infty}^\infty \frac{d\omega}{2\pi}S_F(\omega)|f_{\omega}(t)|^2,
\end{equation}
with 
$m\equiv k_0^2/(D_fv_f)^2\!-\!k_0^2/(D_i D_fv_i^2)$. The nonlinear equation (\ref{fomega_eq}) must be complemented by initial conditions, which we find by ensuring the continuity of the field envelope $\mathcal{E}$ and of its time derivative at the temporal interface. From the prequench solution (\ref{phi_prequench}), these conditions yield
\begin{equation}
\label{eq:init}
f_\omega(0)=1,\ \ \dot{f_\omega}(0)=-i(\omega+\mu_i-\mu_f),
\end{equation}
where the factor $\mu_i-\mu_f=k_0/(D_i v_i)\!-\!k_0/(D_f v_f)$ stems from the phase relating the variables $\phi$ and $\mathcal{E}$, see Eq. (\ref{phi_vs_E}), which  changes from $\mu_i t$ to $\mu_f t$ when the quench is performed.

In statistical physics, equations of the form of (\ref{fomega_eq}) have been studied in the context of quenches in the $O(N)$ model, for which they constitute the exact solution in the limit $N\to\infty$ \cite{Chandran2013, Sciolla2013, Smacchia2015, Chiocchetta2017, Maraga2015, Mitra2015, Mitra2016}. In that framework, $m$ is the ``mass parameter'' of the free theory ($g=0$), which becomes self-consistently renormalized to  $m_\text{eff}(t)$ when $g$ is nonzero.  
In the present optical problem, $m_\text{eff}(t)$ is a crucial quantity for the dynamics, in turn related to the total wave intensity after the quench through:
\begin{equation}
\langle|\phi(\br_\perp,z,t)|^2\rangle\!=\!\int \frac{d\omega}{2\pi}S_F(\omega)|f_{\omega}(t)|^2\!=\!\frac{m_\text{eff}(t)\!-\! m}{2gI/D_f},
\end{equation}
where the first equality follows from Eq. (\ref{eq:Ansatz}), using Eqs. (\ref{correlator_q}) and (\ref{correlator_omega}). 
A core property of the equation of motion (\ref{fomega_eq}) is that if $m$ is chosen  negative (which is realized for $D_fv_f^2>D_iv_i^2$), the effective mass $m_\text{eff}(t)$ may vanish at long time and induce a dynamical phase transition \cite{Chandran2013}. The precise condition for this to happen is discussed in the next subsection.

It is interesting to notice, finally, that the initial transverse fluctuations have been completely traced out  in the derivation of Eq. (\ref{fomega_eq}). This means that their precise properties are of no importance for the postquench dynamics, which is solely governed by the frequency fluctuations. This decoupling originates from the factorization  of spatial and frequency fluctuations that we assumed for the incoming wave, Eq. (\ref{eq:spectrum}). As a matter of fact, the essential role of  the transverse fluctuations is here to guarantee a vanishing mean field for the prequench state, Eq. (\ref{eq:mean_field}).

\subsection{Dynamical phase transition}
\label{Sec:DPT}

In \cite{Sotiriadis2010}, it was shown that in nonlinear equations of the type of Eq. (\ref{fomega_eq}), the effective mass $m_\text{eff}(t)$ generically converges at long time to a constant, positive value $m_\text{eff}(\infty)$. A dynamical phase transition then exists if $m_\text{eff}(\infty)$ vanishes for a certain critical (negative) value $m_c$ of the quench parameter $m$. 
To find out whether such a DPT is present in the model  (\ref{fomega_eq},\ref{meff_def}), we use an Ansatz originally proposed in \cite{Sotiriadis2010, Smacchia2015} for calculating $m_\text{eff}(\infty)$: we replace the stationary value of $|f_\omega(t)|^2$ in the r.h.s. of Eq. (\ref{meff_def}) by the result of the free theory ($g=0$) self-consistently evaluated at $m=m_\text{eff}(\infty)$. Since the solution of the free theory is 
\begin{align}
f^\text{free}_{\omega}(t)&=\cos(t\sqrt{w^2D_i/D_f\!+\! m})\\
&-\frac{i(\omega\!+\!\mu_i\!-\!\mu_f)}{\sqrt{\omega^2D_i/D_f\!+\! m}}\sin(t\sqrt{w^2D_i/D_f\!+\! m}),\nonumber
\end{align}
this Ansatz leads to 
\begin{align}
m_\text{eff}(\infty)&= m+\frac{2 g I}{D_f}\int \frac{d\omega}{2\pi}S_F(\omega)\nonumber\\
&\times\frac{(\omega\!+\!\mu_i-\mu_f)^2\!+\!\omega^2D_i/D_f\!+\!m_\text{eff}(\infty)}{2[\omega^2 D_i/D_f\!+\!m_\text{eff}(\infty)]}.
\label{eq:m_infty}
\end{align}
A DPT, if it exists, corresponds to a critical value $ m= m_c$ for which $m_\text{eff}(\infty)=0$. This imposes that
\begin{equation}
\label{eq:mc}
m_c\!=\!-\frac{g I}{D_i}\int \frac{d\omega}{2\pi}S_F(\omega)
\frac{(\omega\!+\!\mu_i\!-\!\mu_f)^2\!+\!\omega^2D_i/D_f}{\omega^2}.
\end{equation}
In this relation, the frequency integral in the r.h.s. is non-divergent at $\omega\to0$ only when  $\mu_i=\mu_f$. This is the ``phase-matching'' condition required for the dispersive medium to display a DPT after the quench, which we assume to be fulfilled from now on:
\begin{equation}
\label{eq:phase_matching}
\ \mu_i=\mu_f\equiv \mu \Leftrightarrow D_i v_i=D_f v_f.
\end{equation}

The phase-matching condition allows us to rewrite the equation of motion (\ref{fomega_eq}) under the following dimensionless form:
\begin{align}
\label{fomega_eq_dimensionless}
&\ddot f_\omega (t)\!+\!\Big[\frac{v_f}{v_i}\omega^2\!+\!m_\text{eff}(t)\Big]f_\omega(t)=0,\\
& m_\text{eff}(t)=  m+\frac{\lambda}{2}\int_{-\infty}^\infty\! \frac{d\omega}{2\pi}
\frac{2\gamma}{(\omega-1)^2+\gamma^2}|f_{\omega}(t)|^2,\nonumber
\end{align}
where we introduced $\lambda\equiv4gI/(D_f\mu^2)$. The mass parameter is now $m\equiv 1-v_f/v_i$, and time and frequencies are expressed in units of $\mu^{-1}$ and $\mu$, respectively. Note that this equation of motion
 depends on  three independent parameters only: the nonlinear coupling strength $\lambda$, the spectral width $\gamma$, and the ratio $v_f/v_i$ of group velocities before and after the quench, this ratio being the natural control parameter allowing to explore the DPT. Equation (\ref{fomega_eq_dimensionless}) is also the form that will be used in our numerical simulations below. At the phase-matching condition (\ref{eq:phase_matching}), the critical quench (\ref{eq:mc}) at which the DPT occurs can be rewritten in terms of a critical value $\left.v_f/v_i\right|_c$ :
\begin{equation}
m_c\equiv 1-\left.\frac{v_f}{v_i}\right|_c=-\frac{\lambda}{4}\frac{1\!+\!\left.v_f/v_i\right|_c}{\left.v_f/v_i\right|_c},
\end{equation}
which yields the critical ratio of group velocities:
\begin{equation}
\label{eq:criticalquench}
\left.\frac{v_f}{v_i}\right|_c=\frac{1}{8}\left[4+\lambda+\sqrt{16+\lambda(24+\lambda)}\right].
\end{equation}
In the next two sections, we characterize the general properties of the solutions of Eq. (\ref{fomega_eq_dimensionless}) in the vicinity of the critical quench, i.e., for $m>m_c$ and $m<m_c$, corresponding to $v_f/v_i<v_f/v_i|_c$  and $v_f/v_i>v_f/v_i|_c$, respectively.  The practical realization of the critical ratio $(\ref{eq:criticalquench})$ for a concrete example of dispersive medium will be discussed in Sec. \ref{Sec:vapor}.

\section{Quench above the critical point}
\label{Sec:above}

\subsection{Divergence of the correlation length}
\label{Sec:xi}

\begin{figure}[h]
\includegraphics[scale=0.3]{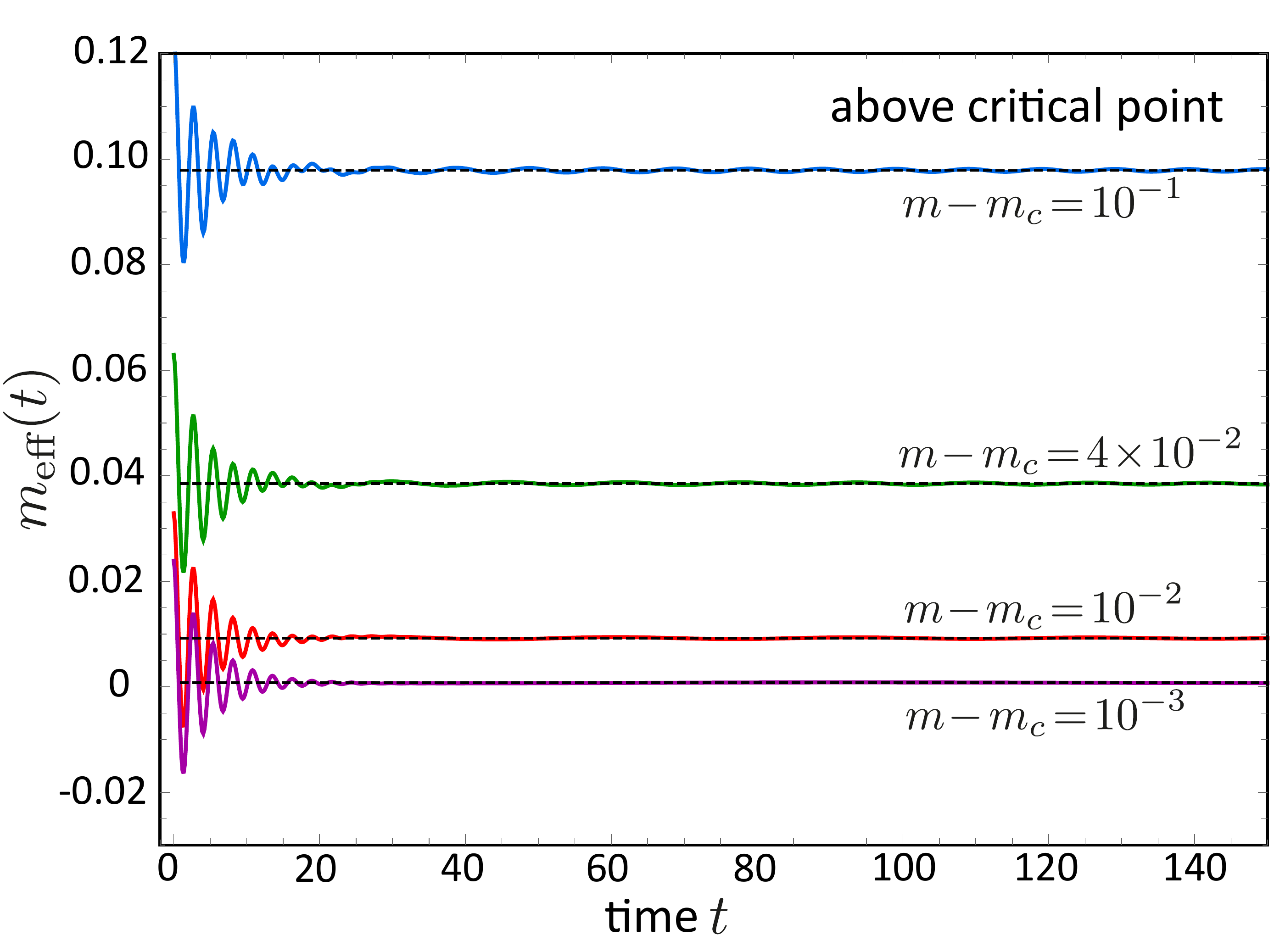}
\caption{Effective mass $m_\text{eff}(t)$ vs. time for a few values of the quench parameter $m$ above its critical value $m_c$, for $\lambda=0.5$ and $\gamma=0.1$.
At long time  $m_\text{eff}(t)$ converges to a finite positive value  $m_\text{eff}(\infty)$. The latter is well captured by the solution of Eq. (\ref{eq:m_infty}), shown as black dashed lines. Here time and frequency are in units of $\mu^{-1}$ and $\mu$, respectively, and $m$, $m_c$ and $m_\text{eff}(t)$ are in units of $\mu^2$.
}
\label{fig:reff_aboverc}
\end{figure}

We first show in Fig. \ref{fig:reff_aboverc} the effective mass $m_\text{eff}(t)$ as a function of time, obtained by solving Eq. (\ref{fomega_eq_dimensionless}) numerically for a few values of $m$ above $m_c$.
As announced, following the quench the effective mass quickly saturates at a finite value $m_\text{eff}(\infty)$ that gets closer and closer to zero as $m$ approaches $m_c$. The saturation value is well described by the solution of Eq. (\ref{eq:m_infty}), shown as dashed lines.

\begin{figure}
\includegraphics[scale=0.3]{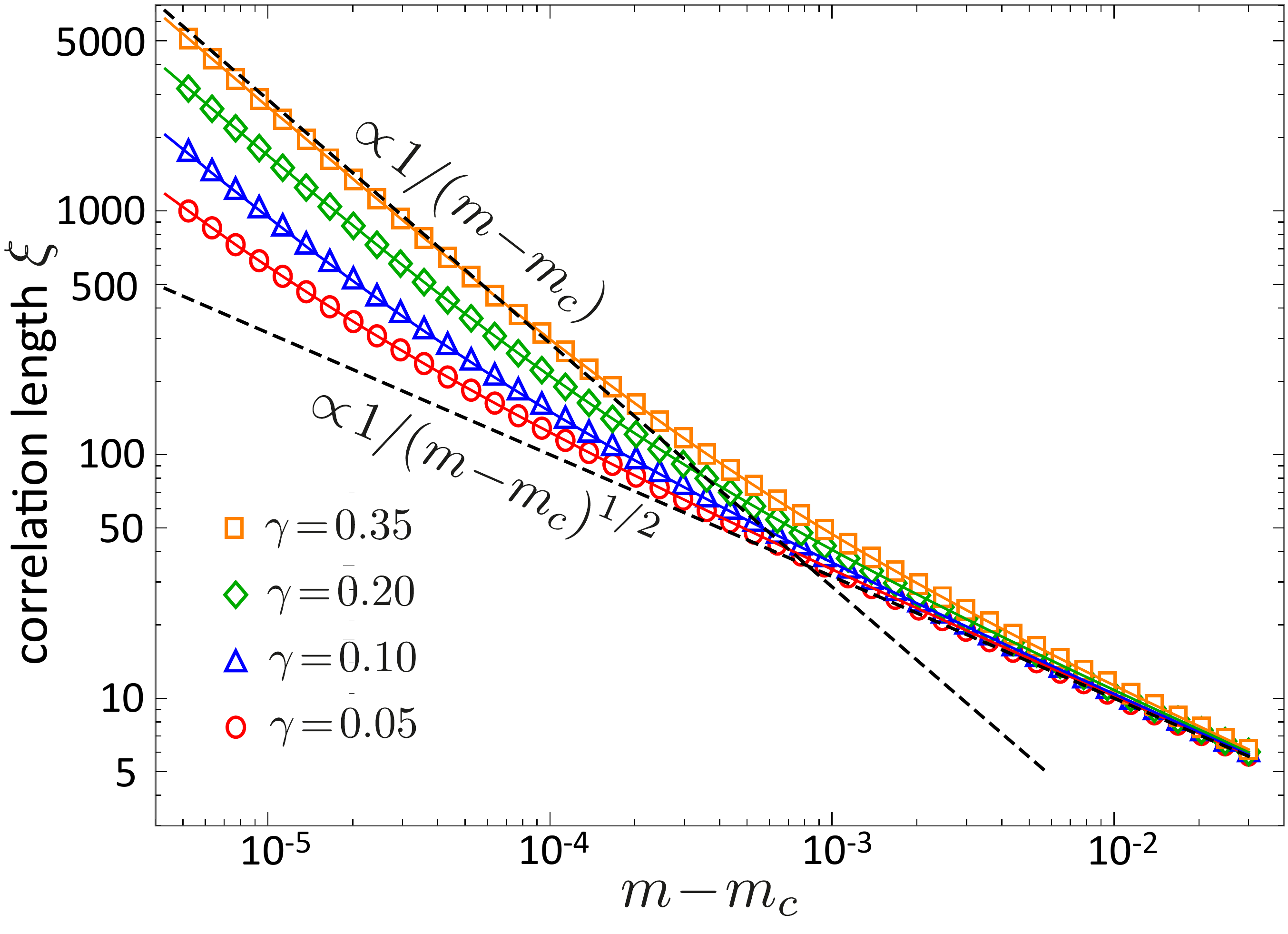}
\caption{Correlation length $\xi\equiv 1/\sqrt{m_\text{eff}(\infty)}$ as a function of the distance $m-m_c$ to the critical point, for different values of the bandwidth $\gamma$. Symbols are numerical results, obtained by solving Eq. (\ref{fomega_eq_dimensionless}). The dashed lines show the asymptotic analytical  predictions  (\ref{xi_nu1}) and (\ref{xi_nu1/2}) in the close vicinity of the DPT and slightly away from it. Here $\lambda=0.5$ and parameters units are the same as in Fig. \ref{fig:reff_aboverc}.
}
\label{fig:xivs_deltar}
\end{figure}

Above the DPT, the saturation value of the effective mass defines a ``correlation length'' $\xi\equiv 1/\sqrt{m_\text{eff}(\infty)}$, which diverges algebraically near the critical point. Fig. \ref{fig:xivs_deltar} shows $\xi$ as a function of the distance $m-m_c$ to the critical point, for different values of the bandwidth $\gamma$ of the frequency spectrum. Whatever $\gamma$ is, we find that $\xi$ diverges algebraically at $m=m_c$, which is a hallmark of the DPT.
The numerical results of Fig. \ref{fig:xivs_deltar}, however, also demonstrate the existence of a \emph{cross-over} between two algebraic scaling laws in the region $0<m-m_c\ll1$. Indeed, above a small but finite value of $m-m_c$ we find $\smash{\xi\sim 1/(m-m_c)^{1/2}}$, whereas below that value one has $\xi\sim1/(m-m_c)$, a scaling which persists up to arbitrarily small $m-m_c$. The existence of this cross-over is a characteristic feature of the  optical DPT. It is, in turn, due to the competition between two antagonistic effects: on the one hand the approach to the DPT, whose critical properties are governed by the infrared frequency limit $\omega\to 0$, and on the other hand the frequency spectrum of the beam, which selects the finite frequency $\omega=1$, see Eq. (\ref{fomega_eq_dimensionless}). This can be explicitly demonstrated  by combining Eqs. (\ref{eq:m_infty}), (\ref{eq:mc}) and (\ref{eq:phase_matching}) so to express $m_\text{eff}(\infty)$  as a function of $m-m_c$:
\begin{equation}
\label{eq:meff_close}
m_\text{eff}(\infty)\!=\!m\!-\!m_c\!-\!\frac{\lambda}{2}\!\int\!\frac{d\omega}{2\pi}\frac{S_F(\omega)m_\text{eff}(\infty)}{2\frac{v_f}{v_i}[\frac{v_f}{v_i}\omega^2+m_\text{eff}(\infty)]}.
\end{equation}
The existence of the cross-over between to critical behaviors becomes clear if one notices that the value of the integral in the r.h.s. depends on which of the functions $S_F(\omega)$ or $[\frac{v_f}{v_i}\omega^2+m_\text{eff}(\infty)]^{-1}$ is the narrowest. At very small $m-m_c$, it is the second one, which selects the infrared frequencies and eventually yields, to leading order in $m_\text{eff}(\infty)\ll1$:
\begin{equation}
\label{xi_nu1}
\xi\equiv\frac{1}{\sqrt{m_\text{eff}(\infty)}}\simeq\frac{1}{m-m_c} \frac{\lambda S_F(0)}{8}\left.\frac{v_i}{v_f}\right|_c^{3/2}.
\end{equation}
On the other hand, at larger $m-m_c$ the frequency spectrum becomes more peaked so that the frequency region around $\omega=1$ becomes the dominant contribution to the integral. As a result, the second term in the r.h.s. of Eq. (\ref{eq:meff_close}) becomes sub-leading, and we find instead:
\begin{equation}
\label{xi_nu1/2}
\xi\simeq\frac{1}{\sqrt{m-m_c}}.
\end{equation}
The two asymptotic laws (\ref{xi_nu1}) and (\ref{xi_nu1/2}) are displayed in Fig. \ref{fig:xivs_deltar} (dashed lines). They match very well  the numerical results without any fit parameter. The cross-over point separating the two regimes  is readily obtained by equating the asymptotes: $m-m_c\simeq \lambda^2\gamma^2/16$. 

Let us finally say a word on the distribution function $f_\omega(t)$ above the critical point. Since the effective mass saturates at long time, $f_\omega(t)$ is approximately given by
\begin{align}
f_{\omega}(t)&\simeq\cos(t\sqrt{\omega^2\!+\! \xi^{-2}})
-\frac{i\omega\sin(t\sqrt{\omega^2\!+\! \xi^{-2}})}{\sqrt{\omega^2\!+\! \xi^{-2}}},
\end{align}
where we used that $v_f/v_i$ is close to one in the vicinity of the critical point.  Coming back to physical problem of light propagation, we see that this solution actually describes the superposition of two optical waves propagating forward and backward in time for $t>0$. This phenomenon, illustrated in Fig. \ref{fig:setup}(b), is characteristic of waves scattered from time-varying dielectric interfaces (see, e.g., \cite{Pendry2022} for a review).

\subsection{Link to the $O(N)$ model and dimensional cross-over}
\label{Sec:ON}

The two asymptotic laws (\ref{xi_nu1}) and (\ref{xi_nu1/2}) can be recast as $\xi\sim (m-m_c)^{-\nu}$, with a critical exponent $\nu$ crossing-over $1$ to $1/2$ when moving away from the transition. To better understand this  cross-over, it is instructive to make contact with the critical properties of the usual  quantum (i.e., zero-temperature) \emph{equilibrium} phase transition of the $\varphi^4$ model with $O(N)$  symmetry that we recall here. 
The $O(N)$ model describes a $N$-component scalar field $\boldsymbol{\Phi}$ in dimension $d$ with Hamiltonian \cite{Moshe2003, Smacchia2015, Chiocchetta2017}
\begin{equation}
\label{Hamiltonia_ON}
H=\int d^dx\frac{1}{2}\left[\boldsymbol{\Pi}^2+(\nabla\boldsymbol{\Phi})^2+m\boldsymbol{\Phi}^2+\frac{u}{12N}(\boldsymbol{\Phi}^2)^2\right],
\end{equation}
where $\boldsymbol{\Pi}$ is the  canonical conjugate momentum of $\boldsymbol{\Phi}$ (for simplicity we use  the same notation $m$  for the mass parameter as in the optical problem). In the limit $N\to\infty$, an Hartree approximation similar to that we used in Sec. \ref{Sec:HFB} allows to exactly map   the Hamiltonian (\ref{Hamiltonia_ON}) onto a quadratic one with effective mass self-consistently given by \cite{Moshe2003, Smacchia2015}
\begin{equation}
m_\text{eff}=m+\frac{u}{12}\int \frac{d^d\bq}{(2\pi)^d}\frac{1}{\sqrt{\bq^2+m}},
\end{equation}
where $|\bq|$ is supposed to be bounded from above by an ultraviolet cutoff. This equation similarly defines an equilibrium quantum critical point $m_c=-(u/12)\int d^d\bq/(2\pi)^d/|\bq|$ where $m_\text{eff}=0$,  showing that a transition only exists when $d>1$. In that case, the correlation length $\xi\equiv \sqrt{m_\text{eff}}^{-1}$ in the vicinity of the critical point obeys the algebraic law $\xi\sim (m-m_c)^{-\nu}$ with
\begin{equation}
\nu=\frac{1}{d-1}\ \,(1\!<\!d\!<\!3),\ \ \ \ 
\nu=\frac{1}{2}\ \, (d\geq 3).
\end{equation}
A comparison with Eq. (\ref{xi_nu1}) suggests that the critical exponent of the optical DPT in dispersive media when $m-m_c\to0$ is the same as the one of second-order, equilibrium quantum  phase transitions in dimension 2.  Furthermore, the smooth change of the critical exponent from $1$ to $1/2$ observed in Fig. \ref{fig:xivs_deltar} can be seen as a \emph{dimensional cross-over}, where the class of the optical DPT turns from that of a quantum phase transition in dimension 2 to that of a quantum phase transition in dimension 3 (the upper critical dimension) as one moves away from the critical point. 
In the recent works \cite{Maraga2015, Chiocchetta2017}, a somewhat related quantum-to-classical cross-over (but at fixed spatial dimension) was reported but, to our knowledge, a dimensional cross-over in a DPT is a novel phenomenon. As mentioned in the previous section, it stems from the peculiar shape of the fluctuation spectrum (\ref{eq:Lorenztian}), which selects out a nonzero frequency when the spectral width $\gamma$ is small enough. Previous works on DPTs in the $O(N)$ models, on the contrary, have been so far restricted to fluctuation spectra centered on $\omega=0$ \cite{Maraga2015}.

\section{Quench below the critical point}
\label{Sec:below}

\subsection{Scale-invariance}

\begin{figure}
\includegraphics[scale=0.3]{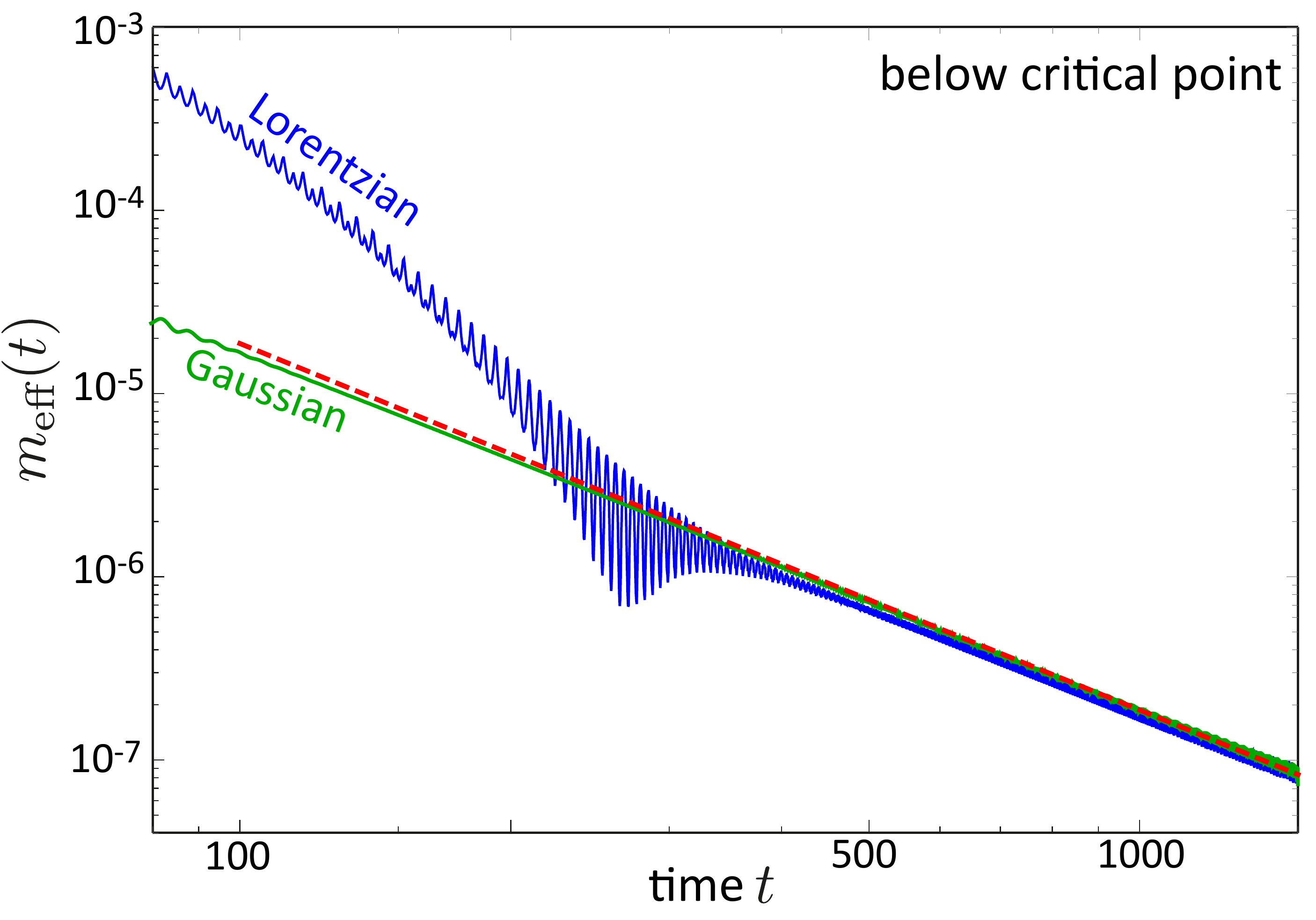}
\caption{\label{fig:mefft_below}
Effective mass $m_\text{eff}(t)$ vs. time  obtained by solving numerically Eq. (\ref{fomega_eq_dimensionless}) for $m-m_c=-0.3$ and $\lambda=4$. The blue and green curves show numerical results for a Lorentzian spectrum (with $\gamma=0.1$) and a Gaussian spectrum (with $\gamma=0.45$), respectively. The dashed red line is the analytical prediction (\ref{meff_long_univ}), consequence of scale invariance below the critical point. Here time and frequency are in units of $\mu^{-1}$ and $\mu$, respectively, and $m$, $m_c$ and $m_\text{eff}(t)$ are in units of $\mu^2$.
}
\end{figure}

Let us now consider quenches below the critical point, namely $m<m_c$ (or $v_f/v_i>v_f/v_i|_c$). Numerical resolution of Eq. (\ref{fomega_eq_dimensionless}) for $m_\text{eff}(t)$ in that case is shown in Fig. \ref{fig:mefft_below}.
After a transient regime, one finds that the effective mass decays algebraically as $m_\text{eff}\sim 1/t^2$. Such a behavior was previously reported in the context of the $O(N)$ model in the limit $N\to\infty$, where it was found to be associated with scale invariance \cite{Maraga2015}. To show this, let us consider a rescaling $(t\to \epsilon t,\omega\to\omega/\epsilon)$ of the time and frequency variables in Eq. (\ref{fomega_eq_dimensionless}). The rescaled equation reads
\begin{equation}
\frac{1}{\epsilon^2}\ddot f_{\omega/\epsilon}(\epsilon t)+\Big[\frac{v_f}{v_i}\frac{\omega^2}{\epsilon^2} +m_\text{eff}(\epsilon t)\Big]f_{\omega/\epsilon}(\epsilon t)=0.
\end{equation}
Comparison with Eq. (\ref{fomega_eq_dimensionless}) immediately shows that scale invariance is achieved when $m_\text{eff}(\epsilon t)=(1/\epsilon^2)m_\text{eff}(t)$, i.e., when $m_\text{eff}(t)\propto 1/t^2$, in agreement with the numerical observation of Fig. \ref{fig:mefft_below}. Scale invariance is thus a core dynamical property of the nonlinear dispersive medium below the critical quench. Below we show how this feature manifests itself in the scaling properties of the  distribution function $f_\omega(t)$.

\subsection{Scaling properties of the distribution function}

The scale invariance property of the equation of motion below the critical point is expected to give rise to a specific, universal self-similar behavior of the distribution function $f_\omega(t)$. To disclose this behavior, we look for asymptotic expressions of the solution of Eq. (\ref{fomega_eq_dimensionless}) in the regime of scale invariance where $m_\text{eff}(t)=a/t^2$, with $a$ a numerical constant to be determined. In the context of the $O(N)$ model, a methodology to achieve this goal was proposed in \cite{Maraga2015} in a particular ``deep quench'' limit where the initial conditions for $f_\omega$ and $\dot{f}_\omega$ are constant. This is not the case in the present optical DPT, which requires to adapt the method of \cite{Maraga2015}, as we now discuss.

For $\omega\geq0$, the general solution of  Eq. (\ref{fomega_eq_dimensionless}) with $m_\text{eff}(t)=a/t^2$ is given by
\begin{equation}
\label{gomega_sol}
f_\omega(t)=\sqrt{\omega t}\Big[A_\omega J_\alpha\Big(\sqrt{\frac{v_f}{v_i}}\omega t\Big)+B_\omega J_{-\alpha}\Big(\sqrt{\frac{v_f}{v_i}}\omega t\Big)\Big]
\end{equation}
where $\alpha=\sqrt{1/4-a}$, and  $A_\omega$ and  $B_\omega$ are coefficients  to be found.
 This solution is only expected to hold beyond a certain time scale $t_0$ beyond which scale invariance emerges \cite{footnote}. Therefore, to find $A_\omega$ and $B_\omega$ we cannot directly use the initial conditions for $f_\omega$, but instead we should to match the solution  (\ref{gomega_sol}) with an approximate expression of $f_\omega(t)$ at $t=t_0$ \cite{Maraga2015}. To find the latter, we extrapolate from the initial condition using the Taylor expansion $f_\omega(t_0)=f_\omega(0)+t_0 \dot{f}_\omega(0)+O(t_0^2)$. From Eq. (\ref{eq:init}), we infer
\begin{equation}
\label{eq:fomegat0}
f_\omega(t_0)\simeq 1-i\omega t_0+O(t_0^2).
\end{equation}
On the other hand, Eq. (\ref{gomega_sol}) in the limit $\omega t_0\ll1$ provides
\begin{align}
f_\omega(t_0)&\simeq
 A_\omega(\omega t_0)^{\alpha+1/2}\frac{(1/2\sqrt{v_f/v_i})^\alpha}{\Gamma(\alpha+1)}\nonumber \\
&\label{eq2:fomegat0}+B_\omega(\omega t_0)^{-\alpha+1/2}\frac{(1/2\sqrt{v_f/v_i})^{-\alpha}}{\Gamma(-\alpha+1)}.
\end{align}
In this expression, one can show that the scaling $(\omega t_0)^{\alpha+1/2}$ of the first term in the r.h.s. eventually provides values of $\alpha$ which are incompatible with the self-consistent relation obeyed by $m_\text{eff}(t)$ [Eq. (\ref{fomega_eq_dimensionless})]. This imposes that $A_\omega=0$. Comparing Eq. (\ref{eq:fomegat0}) and (\ref{eq2:fomegat0}) then implies that $B_\omega=B(\omega t_0)^{\alpha-1/2}(1+B'\omega t_0/2)$, with $B$ and $B'$ prefactors independent of $\omega$, such  that:
\begin{equation}
\label{gomega_sol2}
|f_\omega(t)|^2\simeq |B|^2\frac{t}{t_0} (\omega t_0)^{2\alpha}(1\! +\! B'\omega t_0)  J_{-\alpha}^2\Big(\sqrt{\frac{v_f}{v_i}}\omega t\Big).
\end{equation}
The last step of the calculation consists in inserting this result into the self-consistent equation (\ref{fomega_eq_dimensionless})  obeyed by the effective mass $m_\text{eff}(t)$. Matching both sides  of this equation in the long-time limit imposes the value of $a$. The details of this procedure are presented in Appendix \ref{Appendix} for clarity. It leads to the only possible value $\alpha=1/4$, equivalently $a=3/16$, so that:
\begin{equation}
\label{meff_long_univ}
m_\text{eff}(t)=\frac{3}{16}\frac{1}{t^2}
\end{equation}
for quenches below the critical point.
As shown in Fig. \ref{fig:mefft_below}, this analytical result captures very well the numerical simulations at long time with, in particular, the correct prefactor $3/16$. 

It should be noted that, according to the above analysis,  \emph{both} the scaling law $\sim 1/t^2$ and the  prefactor $3/16$ follow from the property of scale invariance and are, in this sense, universal. In particular, the law (\ref{meff_long_univ}) a priori holds for different types of frequency spectra. We have verified this by numerically computing $m_\text{eff}(t)$ for a Gaussian spectrum $S_F(\omega)=(\sqrt{2\pi}/\gamma)\exp[-(\omega-\mu)/(2\gamma)^2]$. The result, shown in Fig. \ref{fig:mefft_below}, converges as well  to the prediction (\ref{meff_long_univ}) (the time scale $t_0$ is even faster than for the Lorentzian spectrum in that case).

The value $\alpha=1/4$ also governs the asymptotic, frequency-time scaling of the distribution function below the critical quench, which follows from Eq. (\ref{gomega_sol2}):
\begin{numcases}{|f_\omega(t)|^2\sim} 
(\Lambda t)^{1/2} &$ \omega t\ll1$ \label{scaling_fomega1}\\
\Big( \frac{\Lambda}{\omega}\Big)^{1/2} \cos^2(\omega t-\frac{\pi}{8})& $\omega t \gg 1$ \label{scaling_fomega2}.
\end{numcases} 
\begin{figure}
\includegraphics[scale=0.8]{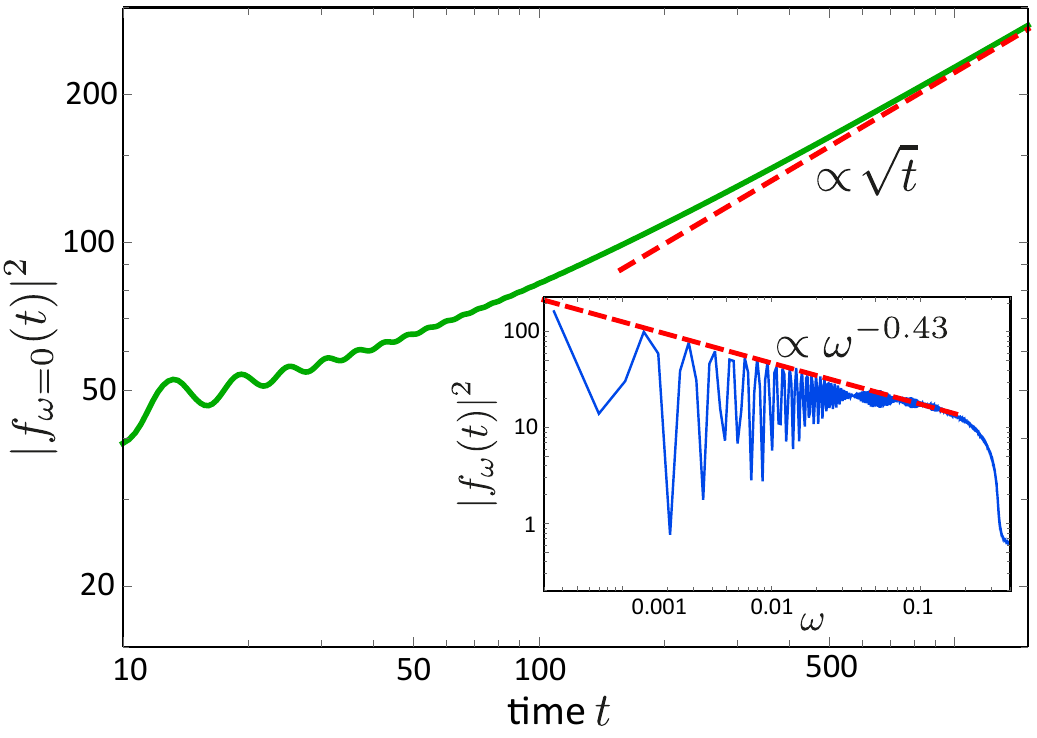}
\caption{
Main plot: distribution function $|f_\omega(t)|^2$ at zero frequency as a function of time, obtained by solving numerically Eq. (\ref{fomega_eq_dimensionless}) for $m-m_c=-0.3$, $\lambda=4$ and $\gamma=0.1$. At long time $|f_0(t)|^2\propto \sqrt{t}$, in agreement with Eq. (\ref{scaling_fomega1}). 
Inset: distribution function $|f_\omega(t)|^2$ as a function of $\omega$. Here time are averaged over a small temporal window of width $\Delta t=54$, centered around $t=1420$. The numerics suggests $|f_\omega(t)|^2\propto \omega^{-0.43}$ at small frequency, close to the prediction (\ref{scaling_fomega2}).  Parameters units are the same as in Fig. \ref{fig:reff_aboverc}.
}
\label{fig:g2_ell_growth}
\end{figure}
To verify this analysis, in Fig. \ref{fig:g2_ell_growth} we show the distribution $|f_\omega(t)|^2$ numerically obtained from Eq. (\ref{fomega_eq_dimensionless}). The main plot shows the $|f_0(t)|^2$, which indeed scales like $t^{1/2}$ at long time, in agreement with Eq. (\ref{scaling_fomega1}). The inset also shows $|f_\omega(t)|^2$ as a function of frequency for a fixed long time. It suggests $|f_\omega(t)|^2\sim \omega^{-0.43}$, close to the prediction (\ref{scaling_fomega2}) [the small deviation from $\omega^{-1/2}$ observed here is due to finite-time effects, Eq. (\ref{scaling_fomega2}) stricty holding  in the limits $\omega t_0\ll1$, $t/t_0\gg1$ and $\omega t\gg1$].
\begin{figure}
\includegraphics[scale=0.3]{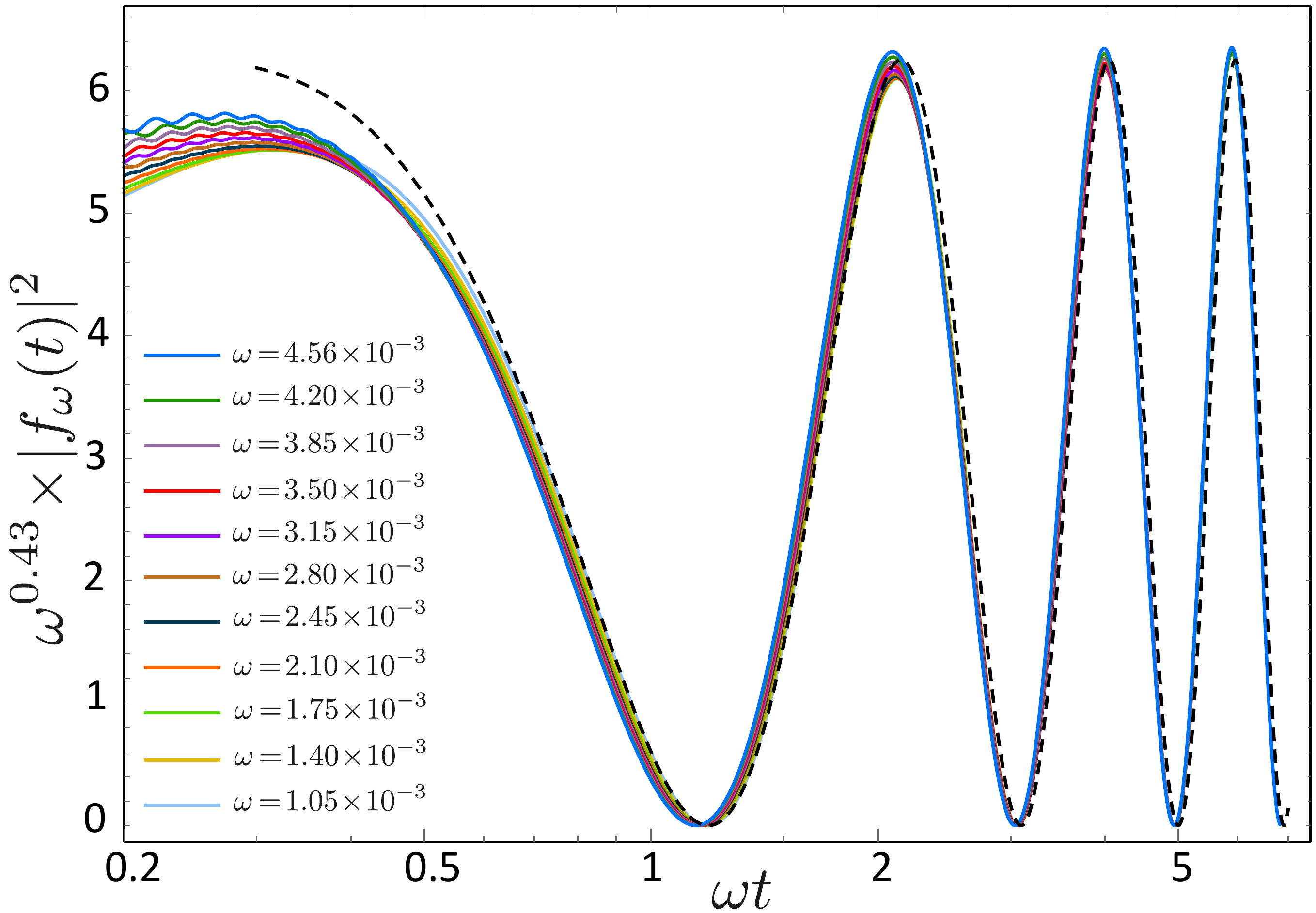}
\caption{
Rescaled distribution $\omega^{0.43} |f_\omega(t)|^2$ as a function of the product $\omega t$ for several values of $\omega$, obtained by solving numerically Eq. (\ref{fomega_eq_dimensionless}) for $m-m_c=-0.3$, $\lambda=4$ and $\gamma=0.1$. The dashed curve is the theoretical prediction (\ref{scaling_fomega2}), in which the only free parameter is the (nonuniversal) prefactor.
 Parameters units are the same as in Fig. \ref{fig:reff_aboverc}.}
\label{fig:fomega_cos}
\end{figure}
In Fig. \ref{fig:fomega_cos}, finally, we plot  the numerical distribution $\omega^{0.43} |f_\omega(t)|^2$ as a function of the product $\omega t$ for several values of $\omega$. The  curves at different frequencies all fall on a single one and oscillate as $\cos^2(\omega t-\pi/8)$, directly confirming the prediction (\ref{scaling_fomega2}). 

As originally pointed out in \cite{Chandran2013, Sciolla2013, Maraga2015} in the context of the $O(N)$ model , both asymptotic relations (\ref{scaling_fomega1}) and (\ref{scaling_fomega2}) can be seen as parts of the general scaling law $|f_\omega(t)|^2=L^{1/2}(t)\mathcal{F}[\omega L(t)]$, which closely ressembles the coarsening dynamics expected when quenching a classical system below a critical point. The quench process then gives rise, for $t>0$, to the local formation of domains of size $L(t)\sim t$ growing linearly in time \cite{Biroli2016}.

\section{Example: light in atomic vapors}
\label{Sec:vapor}

A representative example of nonlinear dispersive medium for light are vapors of hot atoms optically illuminated in the vicinity of an atomic resonance. Recently, this platform has been extensively used to investigate a variety of non-equilibrium phenomena with light \cite{Martone2021, Martone2023, Sun2012, Azam2022, Vocke2015, Santic2018, Fontaine2018, Steinhauer2022, Abuzarli2022, Abobaker2023}. In this section, we discuss under which conditions they could be also exploited to explore the dynamical phase transition studied in the present paper.

Let consider an ensemble of two-level atoms consisting of a ground state $|g\rangle$ and an excited state $|e\rangle$. We denote by $\omega_0$ the resonance frequency between these two states, and by $\Gamma$ the decay rate of the excited state. According to the discussion in Sec. \ref{Sec:DPT}, observing the DPT requires to perform a temporal change of the group velocity $v$ in the vapor from $v_i$ to $v_f$, the critical quench being achieved for the ratio $v_f/v_i=v_f/v_i|_c$ defined by Eq. (\ref{eq:criticalquench}). When $v_f/v_i<v_f/v_i|_c$, the postquench  optical beam lies in the ``normal'' dynamical phase, characterized by a finite effective mass and a finite correlation length, as discussed in Sec. \ref{Sec:above}.  When $v_f/v_i>v_f/v_i|_c$, the beam instead lies in the coarsening phase and the  effective mass decays algebraically, see Sec. \ref{Sec:below}. 

In order to be able to observe the DPT in practice, three constraints must be satisfied: (1) Because $v_f/v_i|_c>1$ [see Eq. (\ref{eq:criticalquench})], crossing the DPT requires the postquench velocity $v_f$ to be larger than the prequench velocity $v_i$, (2) the prequench and postquench dispersion parameters should obey the phase-matching condition (\ref{eq:phase_matching}), and (3) the nonlinear parameter $\lambda$ should be positive. In an atomic vapor, a temporal quench of the dispersion parameters can be achieved by exploiting the dependence of the group velocity $v$ and the quadratic dispersion $D$ upon the detuning $\Delta\equiv\omega-\omega_0$ of the laser exciting the transition: a change from $\Delta_i$ to $\Delta_f$ changes $v_i\equiv v(\Delta_i)$ to $v_f\equiv v(\Delta_f)$, and similarly $D_i\equiv D(\Delta_i)$ to $D_f \equiv D(\Delta_f)$. To express these quantities, we assume that the  vapor is dilute, so that its refractive index $n$ depends on the detuning as $n(\Delta)\equiv 1-(6\pi\rho\Gamma/2k_0^3)\Delta/(\Delta^2+\Gamma^2/4)$, where $\rho$ is the atom density and $k_0\equiv \omega_0/c$ \cite{Morice1995}. The group velocity and the quadratic dispersion parameter respectively follow from $v\equiv (\partial k/\partial\omega)^{-1}$ and $D\equiv (1/2)(\partial^2 k^2/\partial\omega^2)$, where $k\equiv n\omega/c$ is the wave number in the vapor.

\begin{figure}
\includegraphics[scale=0.6]{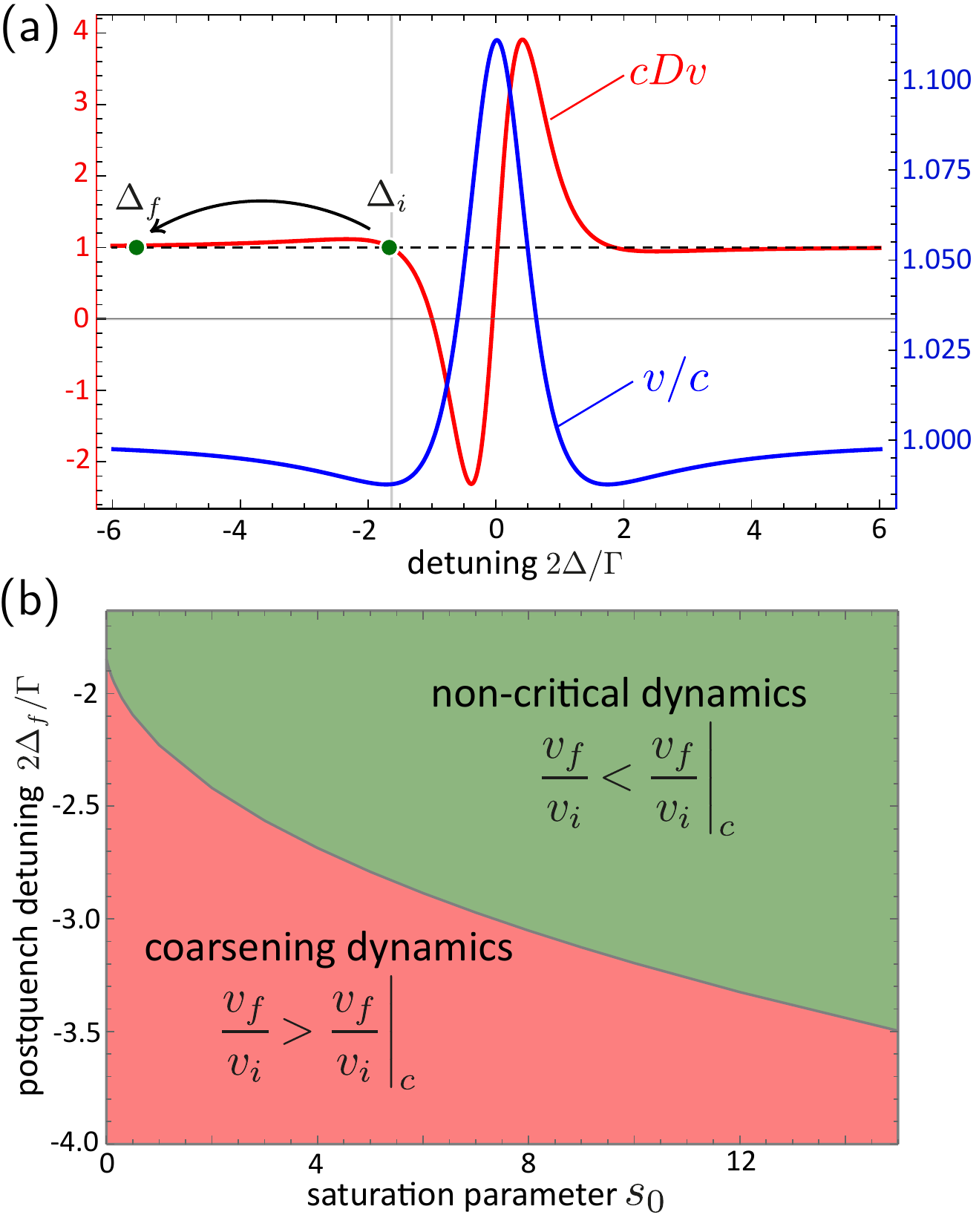}
\caption{
(a) Detuning dependence of the group velocity $v$ and of the product $Dv$ (with $D$ the quadratic dispersion) in a dilute atomic vapor. By choosing $\Delta_i\simeq (-\sqrt{3}+\Gamma/2\omega_0)\Gamma/2$ and quenching to a large negative detuning $\Delta_f$, one simultaneously satisfies the conditions $v_f/v_i>1$, $D_f v_f=D_i v_i$ and $\Delta_f<0$ required to observe the DPT.
The horizontal dashed line is a guide to the eye, identifying the condition $D_f v_f=D_i v_i$.
(b) Phase diagram of the DPT  in the $(s_0,\Delta_f)$ plane, for the choice $\Delta_i\simeq (-\sqrt{3}+\Gamma/2\omega_0)\Gamma/2$.
In both plots, we take $6\pi\rho/k_0^3=10^{-2}$ for the atomic density and $2\omega_0/\Gamma=10$ for the quality factor of the transition.}
\label{fig:spectral_vapor}
\end{figure}

A possible configuration satisfying the above conditions (1), (2) and (3) is illustrated in Fig. \ref{fig:spectral_vapor}(a), where we show the group velocity $v$ and the product $D v$ as a function of $\Delta$: by quenching the detuning from the initial value $\Delta_i\simeq (-\sqrt{3}+\Gamma/2\omega_0)\Gamma/2$ to  a final one $\Delta_f<0$ such that $|\Delta_f|\gg \Gamma/2$, one simultanesously realizes $v_f/v_i>1$, $D_fv_f=D_i v_i$ and $\lambda>0$ [the latter condition follows from the proportionality relation $\lambda\propto g\propto -\Delta_f$ in an atomic vapor, see Eq. (\ref{gI_vapor}) below].

In the configuration described above, the two dynamical phases of the DPT can be probed by varying the ratio $v_f/v_i$ via $\Delta_f$ around the critical value $v_f/v_i|_c$. The latter is identified by the relation  Eq. (\ref{eq:criticalquench}), which is a function of the nonlinear parameter  $\lambda\equiv 4gI D_f v_f^2/k_0^2$. In an ensemble of two-level atoms, the product $gI$ is conveniently expressed in terms of the resonant saturation parameter $s_0$, which is the ratio of the laser intensity to the intensity required to saturate the atomic transition \cite{Silva2023}:
\begin{equation}
\label{gI_vapor}
gI=-\frac{6\pi\rho}{k_0}\frac{\Delta_f \Gamma^3/4}{(\Delta_f^2+\Gamma^2/4)^2} s_0.
\end{equation}
Inserting this relation into Eq. (\ref{eq:criticalquench}) and expliciting the $\Delta_f$ dependence of the ratio $v_f/v_i$, we infer the phase diagram of the DPT in the plane $(s_0,\Delta_f)$ for the atomic vapor, see Fig. \ref{fig:spectral_vapor}(b). For a given laser intensity $s_0$, this diagram indicates that the DPT can be crossed by choosing a large enough (negative) value of the postquench detuning. As the nonlinearity is increased (larger $s_0$), larger detunings are required because of the increase of the critical ratio $v_f/v_i|_c$.

\section{Conclusion}
\label{Sec:conclusion}

In this work, we have provided theoretical evidence for a dynamical phase transition for fluctuating optical beams propagating in nonlinear dispersive media. The existence of this DPT fundamentally relies on a mapping between the nonlinear dispersive wave equation  in the slowly-varying envelope approximation and a massive $\varphi^4$ theory. From this observation, the DPT can be triggered by a temporal change of the dispersion parameters, which simulates a quench in the corresponding $\varphi^4$ model. In this perspective, we have identified the precise phase-matching condition required for the DPT to occur.
Generally speaking, the idea of applying temporal quenches to dielectric media has recently gained more and more interest in optics \cite{Pendry2022}, even though it has so far not been much explored in nonlinear media.

By numerically and theoretically investigating  the optical DPT in the vicinity of the critical point, we have connected its critical exponent to the one of equilibrium quantum phase transitions in the $\varphi^4$ theory. Slightly above the critical point, we have also disclosed a dimensional cross-over of the critical exponent. This cross-over is a characteristic feature of the optical problem, stemming from the peculiar shape of the fluctuation spectrum which competes with the infrared physics of the transition by favoring a finite optical frequency. Below the transition, we have numerically and theoretically described the postquench coarsening dynamics. In particular, because it describes a DPT with quantum-like critical properties, our analytical approach in this regime differs from those of previous work \cite{Maraga2015}, which focused on a classical, ``deep-quench'' limit.

The DPT discussed in this work is an example of fixed point arising in the short-time prethermal regime of a weakly nonlinear  system \cite{Mitra2015} and, in that, does not involve any inelastic scattering processes. Such processes are nevertheless present in the original nonlinear dispersive equation. They are expected to make the system deviate from the fixed point at long time and, eventually, to thermalize it. While the cross-over from prethermalization to thermalization in quantum fluids has been recently studied in a few cases \cite{Buchhold2015, Buchhold2016, Regemortel2018, Duval2023}, its general description in situations where a  prethermal DPT is present remains an open problem.

\begin{acknowledgments}
Financial support from the Agence Nationale de la Recherche (grant ANR-19-CE30-0028-01 CONFOCAL) is gratefully acknowledged. The author is indebted to Quentin Glorieux, Dominique Delande, Maxime Jacquet and Giovanni Martone for helpful discussions.
\end{acknowledgments}

\appendix
\section{Determination of the scaling factor $a$}
\label{Appendix}

To find the value of the scaling factor $a$, we use that, below the critical point, the solutions $m_\text{eff}(t)=a/t^2$ for the effective mass and Eq. $(\ref{gomega_sol2})$ for the distribution function are related through Eq. (\ref{fomega_eq_dimensionless}) at long enough time. Introducing the rescaled time $x\equiv \sqrt{v_f/v_i}t$, Eq. (\ref{fomega_eq_dimensionless}) reads:
\begin{align}
\label{eq:aR(x)}
\frac{v_f}{v_i}\frac{a}{x^2}=m+\Lambda R(x),
\end{align}
where $\Lambda\equiv\lambda|B|^2\sqrt{v_i/v_f}t_0^{2\alpha-1}/2$ (with $\alpha=\sqrt{1/4-a}$) and
\begin{equation}
R(x)\!=\!x
\int_{-\infty}^\infty\!\frac{d\omega}{2\pi}\frac{2\gamma}{(\omega\!-\!1)^2\!+\!\gamma^2}\omega^{2\alpha}(1\!+\!B't_0\omega)J^2_{-\alpha}(\omega x).
\end{equation}
Then we follow the method proposed in \cite{Maraga2015} and expand $R(x)$ at large $x$. The expansion reads:
\begin{equation}
R(x)=C_0(\alpha,B')+\frac{C_1(\alpha)}{x^{2\alpha}}+\frac{C_2(\alpha,B')}{x^{2\alpha+1}}+\frac{C_3(\alpha,B')}{x^{2}}+\ldots
\end{equation}
The coefficients $C_i$ of this expansion all depend on $\alpha$ and $B'$, except $C_1$ (they also all depend on $\gamma$, but this dependence is not relevant in the reasoning). 

In order for Eq. (\ref{eq:aR(x)}) to be satisfied, both prefactors $C_1$ and $C_2$ should be zero. Since $C_1$ is independent of $B'$, the condition $C_1(\alpha)=0$ is the one that restricts the possible values of $\alpha$. From the expansion of $R(x)$ at large $x$ we find: 
\begin{equation}
C_1(\alpha)\propto \frac{1}{\Gamma(1/2-2\alpha)\Gamma(1/2-\alpha)},
\end{equation}
where $\Gamma$ is the gamma function. This leads to the possible sets of values $\alpha=\{1/4+n/2\}$ or $\alpha=\{1/2+n\}$, with $n$ an integer. Among these values, only the one $\alpha=1/2$ is compatible with a positive effective mass, i.e., $a>0$. The condition $C_2(\alpha,B')=0$, on the other hand, enforces the value of $B'$, while the equations $m+\Lambda C_0(\alpha,B')=0$ and $\Lambda C_3(\alpha,B')=(v_f/v_i)a$ fix the values of $B$ and $t_0$.


\begin{thebibliography}{99}
\bibitem{footnote}
 The time scale $t_0$ is not universal and, in particular, depends on the form of the fluctuation spectrum. For the case of Lorentzian spectrum with $\gamma=0.1$ shown in Fig. \ref{fig:mefft_below}, on has, e.g., $t_0 \simeq 300/\mu$. For a Gaussian spectrum, $t_0$ is much shorter.
\end{thebibliography}
\end{document}